\documentclass[10pt,aps,twocolumn,prd,superscriptaddress,noshowpacs,nofootinbib,noshowkeys,floatfix]{revtex4}
\usepackage[dvips]{graphics,graphicx}
\usepackage{amsmath, amssymb}
\usepackage{multirow}
\usepackage{longtable}
\usepackage{color}
\usepackage[normalem]{ulem}  

\renewcommand{\thesection}{\arabic{section}}

\renewcommand{\theequation}{\arabic{section}.\arabic{equation}}

\renewcommand\sout{\bgroup \color{red} \ULdepth=-.5ex \ULset}
\newcommand{\Slash}[1]{\ooalign{\hfil/\hfil\crcr$#1$}}

\begin{document}

\title{Correlations between light and heavy flavors near the chiral crossover}

\author{Chihiro Sasaki}
\affiliation{%
Frankfurt Institute for Advanced Studies,
D-60438 Frankfurt am Main,
Germany
}
\affiliation{%
Institute of Theoretical Physics, University of Wroclaw,
PL-50204 Wroclaw,
Poland
}

\author{Krzysztof Redlich}
\affiliation{%
Institute of Theoretical Physics, University of Wroclaw,
PL-50204 Wroclaw,
Poland
}

\date{\today}

\begin{abstract}
Thermal fluctuations and correlations between the light and heavy-light
mesons are explored within a chiral effective theory implementing heavy
quark symmetry. We show,  that various heavy-light flavor correlations
indicate a remnant of the chiral criticality in a narrow range of temperature  
where the chiral susceptibility exhibits a peak structure.
The onset of the chiral crossover, in the heavy-light flavor correlations,   
is therefore independent from the
light flavors. This indicates that the fluctuations carried by strange
charmed mesons can also be used to identify the chiral crossover,   
which is  dominated by the non-strange light quark dynamics.
\end{abstract}


\maketitle

\section{Introduction}
\label{sec:int}

Modifications in  magnitude of fluctuations for different observables 
are usually considered as an excellent probe of 
a phase transition or its remnant. In heavy-ion collision,
fluctuations related to conserved charges carried by light and strange
quarks play an important role to identify the QCD  chiral crossover 
or deconfinement properties ~ \cite{fluct,R42}.
Recently, however, the  Lattice QCD simulations have revealed,  
that the charmed mesons are deconfined together with light-flavor mesons 
in the  temperature range where the chiral crossover is partly 
restored~\cite{latcharm}. This result indicates that the light-flavor dynamics
interferes non-trivially with the heavy flavors.

In the field theoretical approach,
the physics of heavy-light hadrons is constrained by heavy quark symmetry
which emerges in the heavy quark mass limit~\cite{HQS1,HQS2}.
The pseudo-scalar and vector charmed mesons form the lowest spin multiplets
$H$, and their low-energy dynamics is dominated by interactions with
Nambu-Goldstone bosons associated with spontaneous chiral symmetry
breaking~\cite{Georgi,Wise,BD,YCCLLY}. The chiral partner of $H$ is
embodied as the second-lowest spin multiplets $G$~\cite{NRZ,BH}.
The mass splitting between $G$ and $H$ is proportional to the chiral order
parameter, and its thermal/dense evolution characterizes partial restoration
of the chiral symmetry in a medium~\cite{Friman,HRS,qsr4,charm:cs,nagoya}.

Recently, a self-consistent effective theory implementing the chiral and
heavy quark symmetry has been formulated at finite temperature~\cite{charm:cs}.
In the present paper, we will use this effective theory to study the fluctuations
in various flavor sectors at finite temperature and vanishing chemical potential.
Our special attention will be paid to the properties of  heavy-light mixed 
correlations to be influenced by the underlying heavy quark symmetry in the 
presence of the chiral crossover.
We will show that the onset of the chiral crossover is well identified 
in the heavy-light flavor correlations,  and that it is independent from the
light flavors.

\section{Effective Lagrangian}
\label{sec:eff}

We utilize  the Lagrangian which includes the mesons with the light and heavy
flavors and their couplings.

To quantify the light-flavor dynamics, we introduce the standard linear
sigma model with three flavors. The main building block is the chiral field 
$\Sigma = T^a\Sigma^a = T^a\left(\sigma^a + i\pi^a\right)$, expressed 
as a $3 \times 3$
complex matrix  in  terms of the scalar $\sigma^a$ and the pseudoscalar $\pi^a$
states. The Lagrangian is given by
\begin{eqnarray}
{\mathcal L}_{\rm L}
&=&
\bar{q}\left(i\Slash{\partial} - g T^a\left(
\sigma^a + i\gamma_5\pi^a
\right)\right)q
\nonumber\\
&&
{}+ {\rm tr}\left[\partial_\mu\Sigma^\dagger\cdot\partial^\mu\Sigma\right]
{}- V_{\rm L}(\Sigma)\,,
\end{eqnarray}
with
\begin{eqnarray}
V_{\rm L}
&=&
m^2{\rm tr}\left[\Sigma^\dagger\Sigma\right]
{}+ \lambda_1\left({\rm tr}\left[\Sigma^\dagger\Sigma\right]\right)^2
\nonumber\\
&&
{}+\lambda_2{\rm tr}\left[\left(\Sigma^\dagger\Sigma\right)^2\right]
{}- c\left(\det\Sigma + \det\Sigma^\dagger\right)
\nonumber\\
&&
{}- {\rm tr}\left[h\left(\Sigma + \Sigma^\dagger\right)\right]\,.\label{lag}
\end{eqnarray}
The $U(1)_A$ breaking effects in Eq.~(\ref{lag}) are accommodated  in 
the determinant terms, whereas the last term, proportional to  $h = T^a h^a$,
breaks the chiral symmetry explicitly.

Heavy-light meson fields,  with negative and positive parity,  are
introduced as~\cite{NRZ,BH}
\begin{eqnarray}
H
&=&
\frac{1 + \Slash{v}}{2}\left[
P_\mu^\ast\gamma^\mu + iP\gamma_5
\right]\,,
\\
G
&=&
\frac{1 + \Slash{v}}{2}\left[
-iD_\mu^\ast\gamma^\mu\gamma_5 + D
\right]\,,
\label{GH}
\end{eqnarray}
and chiral eigenstates are given via
\begin{equation}
{\mathcal H}_{L,R}
=
\frac{1}{\sqrt{2}}\left(G \pm iH\gamma_5\right)\,.
\end{equation}
The relevant operators are transformed under the chiral and heavy quark
symmetries as
\begin{eqnarray}
{\mathcal H}_{L,R}
&\to&
S{\mathcal H}_{L,R}g_{L,R}^\dagger\,,
\\
\Sigma
&\to&
g_L\Sigma g_R^\dagger\,,
\end{eqnarray}
with the group elements $g_{L,R} \in SU(3)_{L,R}$
and $S \in SU(2)_{Q=c}$.

To formulate thermodynamics, we employ the mean field approximation.
We also assume that there is the SU(2) isospin symmetry in the up and 
down quark sector.
This leads to $\sigma_0$ and $\sigma_8$ as non-vanishing condensates,
which contain both strange and  non-strange components.
The  pure non-strange and strange parts are obtained through the following
rearrangement,
\begin{equation}
\begin{pmatrix}
\sigma_q\\
\sigma_s
\end{pmatrix}
=
\frac{1}{\sqrt{3}}
\begin{pmatrix}
\sqrt{2} & 1 \\
1 & -\sqrt{2}
\end{pmatrix}
\begin{pmatrix}
\sigma_0 \\
\sigma_8
\end{pmatrix}\,.
\end{equation}
In this base, the effective quark masses read
\begin{equation}
M_q
=
\frac{g}{2}\sigma_q\,,
\quad
M_s = \frac{g}{\sqrt{2}}\sigma_s\,.
\end{equation}
The chiral fields, $\sigma_{q}$  and  $\sigma_{s}$,  are related 
with the weak decay constants of pions and kaons, respectively, 
via the partially conserved axial current (PCAC) hypothesis:
\begin{equation}
\langle\sigma_q\rangle = f_\pi\,,
\quad
\langle\sigma_s\rangle
= \frac{1}{\sqrt{2}}\left(2f_K - f_\pi\right)\,.
\end{equation}

The scalar heavy-light meson states are embedded in the multiplets,
\begin{equation}
D = (D_q, D_q, D_s)\,,
\end{equation}
where due to the isospin symmetry,  $D_u=D_d=D_q$.
The potential of the heavy-light sector is expressed,
in terms of the meson mean fields, as~\cite{charm:cs}
\begin{eqnarray}
V_{\rm HL}
&=&
2\left(m_0 + \frac{1}{4}g_\pi^q\sigma_q\right)D_q^2
{}+ \left(m_0 + \frac{1}{2\sqrt{2}}g_\pi^s\sigma_s\right)D_s^2
\nonumber\\
&&
{}+ k_0\left(2D_q^2 + D_s^2\right)^2
{}+ 2k_q\sigma_q D_q^4 + \sqrt{2}k_s\sigma_s D_s^4\,.
\nonumber\\
\label{hl}
\end{eqnarray}

The complete thermodynamic potential,
\begin{eqnarray}
\Omega
=
 V_{\rm L} + V_{\rm HL}+\Omega_q \,,
\end{eqnarray}
contains the contributions of light and heavy-light mesons, expressed  
through  Eqs.  (\ref{lag}) and  (\ref{hl}), respectively,
as well as, the quarks contribution
\begin{eqnarray}
\Omega_q
&=&
6 T\sum_{f=u,d,s}
\int\frac{d^3p}{(2\pi)^3}
\left[\ln\left(1-n_f\right)
{}+ \ln\left(1-\bar{n}_f\right)\right]\,,
\nonumber\\
\end{eqnarray}
with the Fermi-Dirac distribution functions, 
$n_f, \bar{n}_f = 1/\left(1 + e^{(E_f \mp \mu_f)/T}\right)$, 
and the quasi-quark energies,  $E_f = \sqrt{p^2 + M_f^2}$.

In the following, we consider  thermodynamics at  $\mu_B=0$, and account 
for the charm and strangeness conservation. 
The model parameters are fixed at $T=0$, and are summarized in Tables~\ref{paraL}
and~\ref{paraHL}.
\begin{table*}
\begin{center}
\begin{tabular*}{12cm}{@{\extracolsep{\fill}}ccccccc}
\hline
$c$ [GeV] & $m$ [GeV] & $\lambda_1$ & $\lambda_2$ & $h_q$ [GeV$^3$]
& $h_s$ [GeV$^3$] & $g$ \\
\hline
$4.81$ & $0.495$ & $-5.90$ & $46.48$ & $(0.121)^3$ & $(0.336)^3$
& $6.5$ \\
\hline
\end{tabular*}
\end{center}
\caption{
Set of parameters in the light sector with
$m_\sigma = 400$ MeV~\cite{SW}.
}
\label{paraL}
\end{table*}
%
\begin{table*}
\begin{center}
\begin{tabular*}{12cm}{@{\extracolsep{\fill}}cccccc}
\hline
$m_0$ [GeV] & $g_\pi^q$ & $g_\pi^s$ & $k_0$ [1/GeV$^2$]
& $k_q$ [1/GeV$^3$] & $k_s$ [1/GeV$^3$] \\
\hline
$1.04$ & $3.78$ & $2.61$ & $-(1/0.74)^2$ & $-(1/0.44)^3$
& $-(1/0.53)^3$ \\
\hline
\end{tabular*}
\end{center}
\caption{
Set of parameters in the heavy-light sector~\cite{charm:cs}.
}
\label{paraHL}
\end{table*}

A naive implementation of heavy-light mean fields causes a rather strong 
mixing to the sigma fields, and it weakens the magnitude of the explicit 
chiral symmetry breaking at finite $T$.
However, as discussed in~\cite{charm:cs}, this defect can be avoided  when 
a certain temperature dependence in the Lagrangian parameters is present.
Such intrinsic effects emerge, formally, via integrals of underlying degrees 
of freedom which interact with a heat bath. Alternatively, they can be 
extracted from the chiral condensates obtained in lattice QCD, resulting 
in quenching couplings $g_\pi^s(T)$ and $k_s(T)$~\cite{charm:cs}~\footnote{
The thermodynamic potential yields four coupled gap-equations for the mean 
fields $\sigma_q, \sigma_s, D_q$ and $D_s$. Instead of solving them 
with a given set of the coupling constants, one uses the chiral condensates 
extracted from the lattice results, $\sigma_q$ and $\sigma_s$, in the range of
$0.5 < T/T_{\rm pc} < 1.5$ as input, and solves the gap equations 
for the two mean fields, $D_q$ and $D_s$, and for the two couplings, 
$g_\pi^s$ and $k_s$.
}.
In calculating thermodynamic quantities, non-trivial modifications 
emerge from the temperature-dependent couplings whose
gradients at each temperature are under controlled.
This enables us to perform the calculations in a thermodynamically 
consistent way.

In the light-flavor sector,
we set the sigma meson mass to  $m_\sigma=400$ MeV,  so that the chiral
crossover temperature comes out to be $T_{\rm pc}=154$ MeV,  as obtained
in lattice QCD~\cite{LQCD}.

\section{Correlations between different flavors}
\label{sec:corr}
\subsection{Chiral susceptibilities}

Chiral phase transition is characterized by various fluctuations to which
the order parameter couples. The chiral susceptibility $\hat{\chi}_{\rm ch}$,
is defined, through
the following 2-by-2 matrix~\cite{SFR}
\begin{equation}
\hat{C} =
\begin{pmatrix}
C_{qq} & C_{qs} \\
C_{sq} & C_{ss}
\end{pmatrix}\,,
\quad
C_{ij}=\frac{\partial^2\Omega}{\partial\sigma_i\partial\sigma_j}\,, 
\end{equation}
as
\begin{equation}
\hat{\chi}_{\rm ch} = \hat{C}^{-1}
=
\begin{pmatrix}
\chi_{qq} & \chi_{qs}\\
\chi_{sq} & \chi_{ss}
\end{pmatrix}\,.
\label{chsus}
\end{equation}

At the chiral symmetry restoration temperature, the chiral susceptibility
yields a peak, which is predominantly linked to the inverse of the effective
sigma-meson mass squared $M_\sigma^2$,  which measures the curvature of 
the potential. Consequently,  in the
non-strange sector, the chiral susceptibility is well approximated by
\begin{equation}
\chi_{qq} \sim \frac{\partial^2\Omega}{\partial\sigma_q^2}
\sim M_\sigma^{-2}\,.
\end{equation}

\begin{figure}
\begin{center}
\includegraphics[width=8cm]{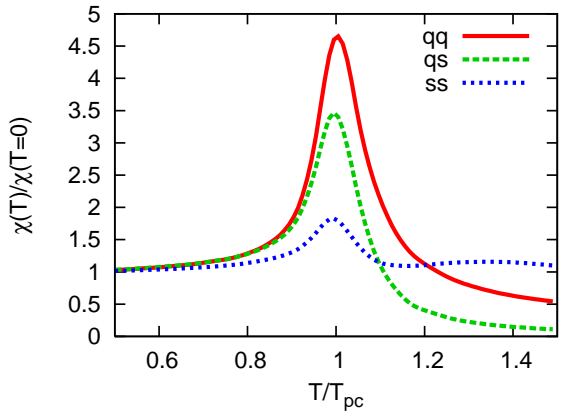}
\caption{
Chiral susceptibilities of $\sigma_q$ and $\sigma_s$.
}
\label{fig:susLL}
\end{center}
\end{figure}

In Fig.~\ref{fig:susLL}, we show the flavor-diagonal  $\chi_{qq}$ and  
$\chi_{ss}$, as well as the
off-diagonal $\chi_{qs}$  chiral susceptibilities.
All these susceptibilities are dominated by an abrupt change of the $\sigma_q$,
leading to a peak around $T_{\rm pc}$. At $T_{\rm pc}$, the strange 
condensate $\sigma_s$, remains substantially larger than $\sigma_q$  
and melts gradually toward higher temperature. This yields a rather broad 
bump in $\chi_{ss}$ around $T \sim 1.35\,T_{\rm pc}$. The flavor-mixed
susceptibility $\chi_{qs}$ is affected strongly by the non-strange dynamics,
and consequently,  any remnant of the modification of $\sigma_s$ is hardly
visible there. Such  properties of $\sigma_s$ can be seen also in a pure 
$SU(2+1)$  flavor   chiral model~\footnote{
  In a conventional treatment with constant couplings, a rate change
  of the chiral order parameter $\sigma_q$ with temperature, 
  is sensitive to the input sigma mass, $m_\sigma$.
  With  increasing $m_\sigma$,  the pseudo-critical temperature
  $T_{\rm pc}$ increases, and the crossover becomes smoother. 
  This suppresses the peaks in
  $\hat{\chi}_{\rm ch}$ around $T_{\rm pc}$. Consequently, for a large 
  $m_\sigma \sim 1$ GeV, the
  $\chi_{ss}$ does not yield a pronounced peak at $T_{\rm pc}$,
  but rather a bump at higher temperature.
}.

The response of the net quark number $n_f$ to the chemical potential
is quantified by the quark number susceptibilities,
\begin{equation}
\chi_f = \frac{\partial n_f}{\partial\mu_f}\,.
\label{qsus}
\end{equation}
In the mean field approximation, the $\chi_{q}$ and $\chi_s$ can be 
expressed as a product of the chiral
condensates and the corresponding susceptibilities as~\cite{SFR}
\begin{equation}
\chi_q \sim \chi_q^{(0)} + \sigma_q^2 \cdot \chi_{qq}\,,
\quad {\rm and } \quad
\chi_s \sim \chi_s^{(0)} + \sigma_s^2 \cdot \chi_{ss}\,,
\end{equation}
where $\chi_{q,s}^{(0)}$ correspond to the non-interacting part.

In the present calculation,  restricted to vanishing chemical
potentials, the peaks of the chiral susceptibilities $\chi_{qq}$ and $\chi_{ss}$
are suppressed by the multiplied condensates $\sigma_{q,s}$,  which monotonically
decrease with temperature. Nevertheless, they
exhibit  sensitivity to the onset of the chiral crossover. At $\mu_f=0$,
the chiral criticality  in the  quark number fluctuations starts to show up 
in  the six-th order cumulant~\cite{FKRS}.
At finite chemical potentials, the $\chi_q$  yields more significance to
the criticality,  carried by the term $\partial\sigma_q/\partial\mu_q$ in
Eq.~(\ref{qsus})~\footnote{
 At zero chemical potential, the coefficient of this term vanishes.
}.

\subsection{Correlations between light and heavy flavors}

To study correlations among the light and heavy-light mesons, we introduce
a matrix in the base of the four mean fields;
\begin{eqnarray}
\hat{\mathcal C}
=
\begin{pmatrix}
{\mathcal C}_{qq} & {\mathcal C}_{qs}
& {\mathcal C}_{qD_q} & {\mathcal C}_{qD_s}\\
{\mathcal C}_{sq} & {\mathcal C}_{ss}
& {\mathcal C}_{sD_q} & {\mathcal C}_{sD_s}\\
{\mathcal C}_{D_qq} & {\mathcal C}_{D_qs}
& {\mathcal C}_{D_qD_q} & {\mathcal C}_{D_qD_s}\\
{\mathcal C}_{D_sq} & {\mathcal C}_{D_ss}
& {\mathcal C}_{D_sD_q} & {\mathcal C}_{D_sD_s}
\end{pmatrix}\,,
\nonumber\\
{\mathcal C}_{ij}
= \frac{\partial^2\Omega}{\partial\sigma_i\partial\sigma_j}\,,
\quad
{\mathcal C}_{iD_j}
= \sqrt{2M_{D_j}}
\frac{\partial^2\Omega}{\partial\sigma_i\partial D_j}\,,
\nonumber\\
{\mathcal C}_{D_iD_j}
= 2\sqrt{M_{D_i}M_{D_j}}
\frac{\partial^2\Omega}{\partial D_i\partial D_j}\,.\label{matrix}
\end{eqnarray}
A set of susceptibilities $\hat{\chi}$ is then defined as
\begin{equation}
\hat{\chi}
= \hat{\mathcal C}^{-1}\,.
\label{HLsus}
\end{equation}
They are associated with the chiral susceptibilities defined in
Eq.~(\ref{chsus}). The explicit relations are summarized in Appendix~\ref{app}.

\begin{figure}
\begin{center}
\includegraphics[width=8cm]{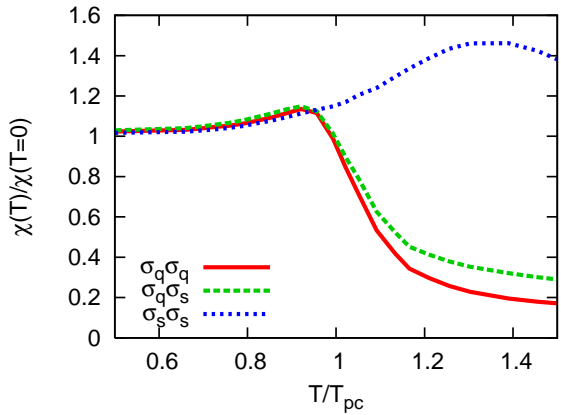}
\caption{
Flavor correlations among the light mesons.
}
\label{fig:susL}
\end{center}
\end{figure}

The light-flavor correlations, obtained from Eq.~(\ref{HLsus}),  
are shown in Fig.~\ref{fig:susL}.
In striking contrast to the chiral
susceptibilities shown in Fig.~1,
they are to a large extent  suppressed, due to the heavy-light mean fields
which act as an extra external source. Nevertheless, the light-flavor correlations
change  their properties around $T_{\rm pc}$. 
The ${qq}$  and ${qs}$ components drop just above  $T_{\rm pc}$, whereas 
the $ss$ component shows a broad bump above $T_{\rm pc}$, similarly  as that
seen  in  $\hat{\chi}_{\rm ch}$ in Fig.~1.

\begin{figure}
\begin{center}
\includegraphics[width=8cm]{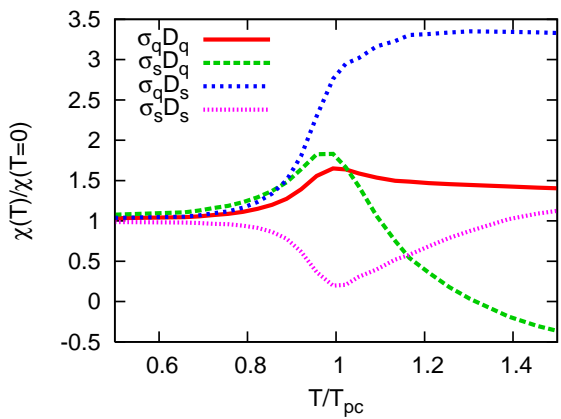}
\caption{
Flavor correlations between the light and heavy-light mesons.
}
\label{fig:susHL}
\end{center}
\end{figure}

\begin{figure}
\begin{center}
\includegraphics[width=8cm]{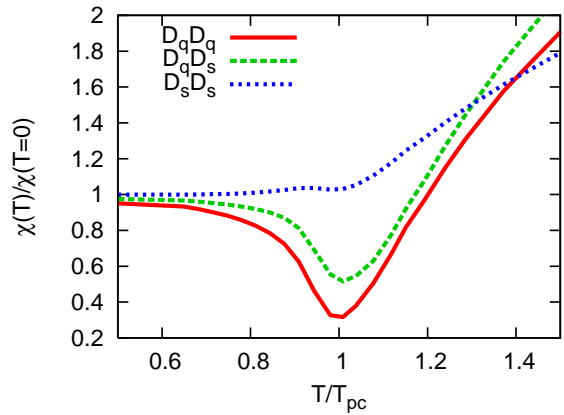}
\caption{
Flavor correlations among the heavy-light mesons.
}
\label{fig:susH}
\end{center}
\end{figure}

Correlations, obtained from  Eqs.~(\ref{matrix}) and (\ref{HLsus}),  
between the light and heavy-light mesons, as well as  those between 
the heavy-light mesons,  are shown  in
Figs.~\ref{fig:susHL} and ~\ref{fig:susH}, respectively.
%

There is a clear influence of remnants of the chiral criticality on  
correlations with heavy flavor. This is the case, 
not only for the susceptibilities involving the non-strange flavors, but also
those with the strange content, which  exhibit a certain qualitative change 
almost in the same temperature range near $T_{\rm pc}$.
This  characteristic feature of the quantities
involving a charm quark, is governed by the
heavy quark symmetry,  which guarantees, in the leading order of the
heavy-quark inverse-mass expansion,  that the physics is independent of
light flavors. The same tendency has also been found in the effective
charmed-meson masses~\cite{charm:cs}. Consequently, the $\chi_{D_iD_j}$
follows the same pattern of a sensitivity to the chiral crossover as
the chiral susceptibilities (see Fig.~\ref{fig:susLL}).
We note that $\chi_{\sigma_sD_q}$ in Fig.~\ref{fig:susHL} changes its sign
at $T/T_{\rm pc} \sim 1.3$, and this does not lead to any instability in the
system since the generalized susceptibility is not precisely the curvature
of the entire potential, see Appendix~\ref{app}.

Certain thermodynamic quantities, to which the chiral condensates couple,
are known to be governed by the underlying $O(4)$ universality around 
$T_{\rm pc}$, even though the chiral symmetry is not exact but violated 
only softly. Just like the chiral symmetry, the heavy quark symmetry is 
also reliable around $T_{\rm pc}$ since the charm quark mass $m_c$ is much 
larger than $T_{\rm pc}$, thus the $1/m_c$ expansion should hold. 
The extended flavor correlations given in Figs.~\ref{fig:susHL} and 
\ref{fig:susH} apparently show that the heavy flavor symmetry constrains 
chiral thermodynamics. 
Further fluctuations beyond the mean field approximation will
modify the flavor correlations on a quantitative level. 
This has been shown on the level of different fluctuations, without heavy flavors,
in terms of
effective chiral models using the functional renormalization method which
accounts for quantum and thermal fluctuations of the chiral fields~\cite{FRG}.
The main modifications due to the quantum fluctuations result in smoothening
susceptibilities across the chiral crossover, however the generic structure
is preserved. Thus, the main feature of the heavy-light correlations 
is also considered to be unaffected.

\section{Conclusions}
\label{sec:conc}

We have studied different correlations between the light and heavy-light flavored
mesons at finite temperature within a chiral effective theory implementing heavy
quark symmetry. A particular attention was attributed to properties of these 
correlations near the chiral crossover temperature, $T_{\rm pc}$.

We have shown that the chiral properties of the correlations  among  the light 
mesons are flavor dependent.
In the $qq$ and $qs$ sector  the qualitative  change of  correlations coincides 
with the chiral crossover.  This is, however, not the case in the $ss$ sector 
where  the modification of the fluctuations  is shifted to a slightly 
higher temperature  above the chiral crossover, due to the large explicit
chiral symmetry breaking for the strange quark.

A striking contrast is found in the correlations involving the heavy-light
meson mean fields. Those fluctuations exhibit certain qualitative changes
around $T_{\rm pc}$, and this feature is independent of the light flavors.
Theoretically, it is anchored to the heavy quark symmetry which guarantees
that the charm quark does not distinguish the non-strange from the strange
flavor in the limit of $m_c \to \infty$.

In the heavy-light system, the heavy quark dynamics is tied to the light
flavor physics, and the thermodynamics is strongly dragged by the chiral
crossover dominated by the non-strange flavors. Consequently, the
fluctuations carried by the strange states can also be used to measure
the onset of the chiral symmetry restoration. The situation is essentially
different from the pure light-flavored system where the observables with
strangeness, e.g. effective masses of the kaon and its chiral partner, 
are rather insensitive to the onset of the chiral crossover.

The lattice QCD studies for the $D_s$ states at finite temperature in
Ref.~\cite{latcharm} strongly suggest that the charmed mesons start to melt at
a temperature close to $T_{\rm pc} \sim 154$ MeV. Thus, the hadronic
picture should not be naively extrapolated to higher temperatures.
However, because of the crossover nature, such dissociation may take place
gradually in some range of temperature. In particular, the correlations
might still be mediated by collective modes like mesonic bound-states 
up to temperatures above the chiral crossover. 
In fact, a study in the $N_f=2$ Nambu-Jona-Lasinio model, that
can handle a bound-state nature of the light mesons and their dissociation,
shows the presence of residual correlations of meson-like states even
above the chiral crossover~\cite{masky}.
This also suggests similar soft modes in the heavy-light sector. 
Therefore, the fluctuations extrapolated slightly 
above $T_{\rm pc}$ may not be totally unrealistic.

An important application is to explore the flavor fluctuations at finite
density and to quantify the chiral modifications of heavy-light hadrons.
There exist a number of works on the charmed mesons in nuclear matter,
in the context of QCD sum rules~\cite{qsr1,qsr2,qsr3,qsr4} and effective
theories~\cite{su4,su42,cc,su43,njl,MGOT,YS,nagoya}. In utilizing our
effective theory, a central task is to introduce a reliable density
dependence of the interaction parameters, which requires a more microscopic
prescription~\cite{SRnew}.

\subsection*{Acknowledgments}

This work has been partly supported by the Hessian LOEWE initiative
through the Helmholtz International Center for FAIR (HIC for FAIR),
and by the Polish Science Foundation (NCN) under
Maestro grant DEC-2013/10/A/ST2/00106.

\appendix

\setcounter{section}{0}
\renewcommand{\thesection}{\Alph{section}}
\setcounter{equation}{0}
\renewcommand{\theequation}{\Alph{section}.\arabic{equation}}

\section{Fluctuations of the light and heavy-light mesons}
\label{app}

In the following, we outline how the chiral susceptibilities are embedded in
the extended flavor susceptibilities~(\ref{HLsus}). We introduce
the following (2x2) block matrices for $\hat{\mathcal C}$:
\begin{equation}
\hat{\mathcal C}
=
\begin{pmatrix}
\hat{\mathcal C}_L & \hat{\mathcal C}_{\rm HL}
\\
\hat{\mathcal C}_{\rm HL} & \hat{\mathcal C}_D
\end{pmatrix}\,,
\end{equation}
with
\begin{eqnarray}
&&
\hat{\mathcal C}_L = \hat{C}\,,
\quad
\hat{\mathcal C}_{\rm HL}
=
\begin{pmatrix}
{\mathcal C}_{qD_q} & {\mathcal C}_{qD_s}
\\
{\mathcal C}_{sD_q} & {\mathcal C}_{sD_s}
\end{pmatrix}\,,
\nonumber\\
&&
\hat{\mathcal C}_D
=
\begin{pmatrix}
{\mathcal C}_{D_qD_q} & {\mathcal C}_{D_qD_s}
\\
{\mathcal C}_{D_sD_q} & {\mathcal C}_{D_sD_s}
\end{pmatrix}\,.
\end{eqnarray}
The heavy-light mixed susceptibilities (\ref{HLsus}) can also be
composed of (2x2)-matrices as
\begin{equation}
\hat{\chi} = \hat{\mathcal C}^{-1}
=
\begin{pmatrix}
\hat{\chi}_{\sigma\sigma} & \hat{\chi}_{\sigma D}
\\
\hat{\chi}_{D\sigma} & \hat{\chi}_{DD}
\end{pmatrix}\,,
\end{equation}
Each matrix is expressed in terms of the chiral susceptibility
$\hat{\chi}_{\rm ch}$ and the matrix $\hat{\mathcal C}_{\rm HL}$ which
represents the curvature of the effective potential in the $\sigma$-$D$
direction. One finds that the following relations hold:
\begin{eqnarray}
\hat{\chi}_{\sigma\sigma}
&=&
\hat{\chi}_{\rm ch} + \hat{\chi}_{\rm ch}\hat{\mathcal C}_{\rm HL}
\hat{\chi}_D\hat{\mathcal C}_{\rm HL}\hat{\chi}_{\rm ch}\,,
\nonumber\\
\hat{\chi}_{\sigma D}
&=&
-\hat{\chi}_{\rm ch}\hat{\mathcal C}_{\rm HL}\hat{\chi}_D\,,
\nonumber\\
\hat{\chi}_{D\sigma}
&=&
-\hat{\chi}_D\hat{\mathcal C}_{\rm HL}\hat{\chi}_{\rm ch}\,,
\nonumber\\
\hat{\chi}_{DD}
&=&
\hat{\mathcal C}_D - \hat{\mathcal C}_{\rm HL}\hat{\chi}_{\rm ch}
\hat{\mathcal C}_{\rm HL}
\equiv
\hat{\chi}_D\,.
\end{eqnarray}


\end{document}